\documentclass[journal]{IEEEtran}
\usepackage{color}
\usepackage{cite}
\usepackage{graphicx}
\usepackage{comment}
\usepackage{subfigure}  
\ifCLASSINFOpdf
 \usepackage[pdftex]{graphicx}
\else
\fi
\begin{document}
\title{Integrated Sensing, Communication, Computing, and Control Meets UAV Swarms in 6G}
\author{Yiyan~Ma,~\IEEEmembership{Member,~IEEE},
        ~Bo~Ai,~\IEEEmembership{Fellow,~IEEE},
        ~Jingli~Li,~\IEEEmembership{Student Member,~IEEE},
        ~Weijie~Yuan,~\IEEEmembership{Member,~IEEE},
        ~Boxiang~He,~\IEEEmembership{Member,~IEEE},
        ~Weiyang~Feng,~\IEEEmembership{Member,~IEEE},
        ~Zhengyu~Zhang,~\IEEEmembership{Member,~IEEE},
        ~Qingqing~Cheng,~\IEEEmembership{Member,~IEEE},
        and~Zhangdui~Zhong,~\IEEEmembership{Fellow,~IEEE}
\thanks{Y. Ma, B. Ai, J. Li, Z. Zhang, and Z. Zhong are with the School of Electronic and Information Engineering, Beijing Jiaotong University, Beijing 100044, China. W. Yuan is with the School of System Design and Intelligent Manufacturing, Southern University of Science and Technology, Shenzhen 518055, China. B. He is with the College of Electronic Science and Technology, National University of Defense Technology, Changsha 410073, China. W. Feng is with the College of Information Science and Engineering, Linyi University, Linyi 276000, China. Q. Cheng is with the School of Electrical Engineering and Robotics, Queensland University of Technology, Brisbane, QLD 4000, Australia. (Corresponding authors: boai@bjtu.edu.cn; jinglili@bjtu.edu.cn).}}
\maketitle
\begin{abstract}
To develop the low-altitude economy, the establishment of the low-altitude wireless network (LAWN) is the first priority. As the number of unmanned aerial vehicles (UAVs) increases, how to support the reliable flying and effective functioning of UAV swarms is challenging. Recently, the integrated sensing, communication, computing, and control (ISCCC) strategy was designed, which could act as effective physical {\it reflex arc} links in the intelligent LAWN system. Thus, in this article, we outline the challenges and opportunities when ISCCC meets UAV swarm in LAWN in 6G. First, we propose a three-layer ISCCC structure for the UAV swarm, which is categorized according to the UAV swarm's requirements, i.e., flying, self-organizing, and functioning. Second, for different scenarios, we study the basic problem, promising technologies, and challenges to design ISCCC for UAV swarms. Third, through a case study that minimizes the flying trajectory error of the UAV swarm based on the ISCCC principle, we show that ISCCC is promising to simultaneously improve the reliability and efficiency of LAWN via jointly designing four components. Finally, we discuss the promising directions for the ISCCC-based UAV swarm in LAWN.
\end{abstract}
\begin{IEEEkeywords}
Integrated Sensing, Communication, Computing, and Control (ISCCC); UAV Swarms; Low-Altitude Wireless Network (LAWN).
\end{IEEEkeywords}
\IEEEpeerreviewmaketitle
\section{Introduction}
As specified by the International Telecommunication Union (ITU), integrated sensing and communication (ISAC) will be a native capability of 6G communication systems \cite{ISAC1}. Compared to 4G and the incipient 5G systems, which are solely designed for information transmissions, ISAC can enhance the sensing capabilities of communication systems and further improve communication performance based on sensing. However, only sensing and communication abilities are still not enough to build an intelligent communication system. Recently, integrated sensing, communication, computing, and control (ISCCC) has been designed in industrial scenarios \cite{ISCCC1}, promising the realization of the closed-loop reaction of the communication systems to the physical situations. Notably, ISCCC will compose an action loop similar to the biological {\it reflex arc}. In addition to ISAC, computing is closely related to signal processing and advanced artificial intelligence (AI) inference, and control is closely related to the terminal situations. Similar to ISAC, the ISCCC could jointly design four components with the scarce channel resources and certain delay limitations, and thus, improve the system efficiency and reliability. Therefore, ISCCC is expected to be the evolution of ISAC for the communication system in 6G and beyond.\par 

Meanwhile, the low-altitude economy has been recognized as one of the new economic growth poles in China. Therein, the low-altitude wireless network (LAWN) is emerging to support the low-altitude flying of diverse unmanned aerial vehicles (UAVs) \cite{LAWN1}. Notably, different from the traditional single UAV-supported communications, the LAWN is expected to support UAV swarms, where plenty of UAVs will perform emerging businesses such as logistics, tourism, transportation management, agricultural production, and emergency communications. To this end, the LAWN faces novel challenges to realize reliable flying and functioning of UAV swarms, including the sense and control of UAV swarms, computation resources allocation among network entities, and cooperative mechanisms between UAV swarms. \par

When ISCCC meets UAV swarms in LAWN, there are kinds of opportunities that emerge. On one hand, ISCCC is promising to provide technical support for UAV swarm flying. Specifically, sensing capability could enable timely perception of swarm situations, and control could maintain the stability of the system operation. On the other hand, ISCCC could improve the efficiency of UAV swarm functioning and self-organizing. Therein, the UAV swarm could turn into an aerial robot based on ISCCC link connections, promising the application of advanced AI models. Meanwhile, based on the wide-range deployment, the UAV swarm could attain high spatial diversity and sensing resolution via ISCCC connections.

In the field of ISCCC designs for UAV systems, there has been some initial exploration. \cite{ISCCC2} introduces a closed-loop ISCCC framework for the satellite-UAV networks and achieves improved control performance by holistically optimizing matrices. \cite{ISCCC3} studies a multi-UAV cooperative network for simultaneous robot control and target localization, which achieves an optimized trade-off between control stability and sensing accuracy via an alternating optimization algorithm. Besides, \cite{ISCCC4} designs a UAV control system assisted by a sensing base station, which applies an active inference framework for joint state estimation, control, and sensing resource allocation.\par

\begin{figure}[t]
    \centering
    \includegraphics[width=3.4in]{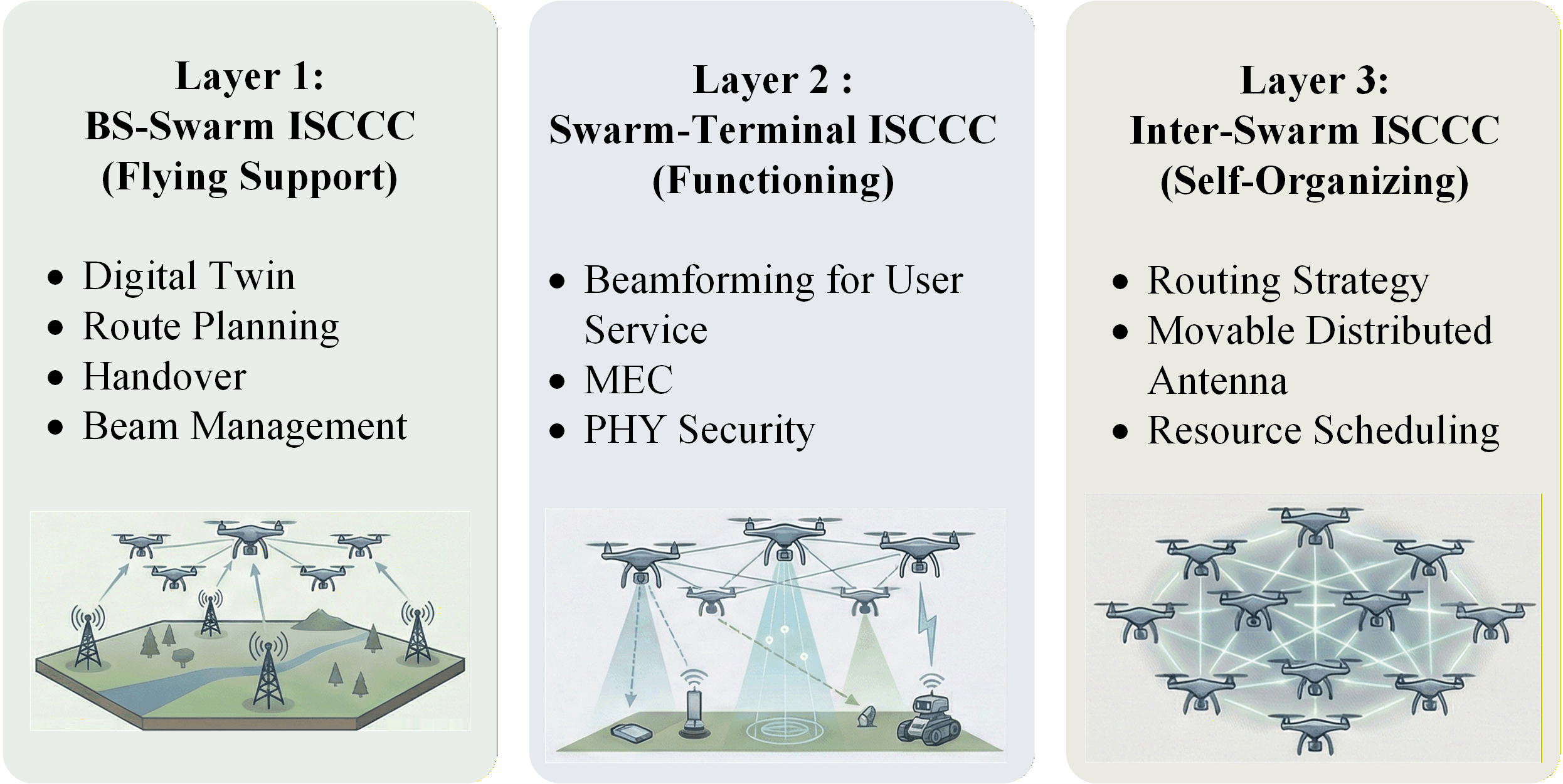}
    \caption{Three-Layer ISCCC structure for UAV swarms in LAWN.}
    \label{fig:1}
    \vspace{-1em}
\end{figure}

Nevertheless, there are various problems that remain to be studied. Typically, the ISCCC-enabled UAV swarms flying, functioning, and self-organizing scenarios and key technologies have not been systematically studied. Therefore, in this article, we investigate the trend of integrating ISCCC to UAV swarms. The main contributions include:
\begin{itemize}
    \item We propose a three-layer ISCCC framework for UAV swarm in LAWN. As shown in Fig. \ref{fig:1}, in the first layer, the UAV swarm flying supported by terrestrial BSs via the ISCCC links is considered, which is further divided into the single base station (BS) and multiple BSs scenarios. In the second layer, the ISCCC links between the UAV swarm and terrestrial terminals are considered, where the UAV swarm acts as arial BS or relays. In the third layer, the design of inter-swarm ISCCC links for swarm self-organizing is discussed, which emphasizes on the cooperative multiple antenna system designs.  
    \item We study the design principles, basic challenges, and promising solutions for ISCCC designs in different layers. In the BS to UAV swarm layer, the basic ISCCC design principle is introduced, and ISCCC designs considering digital twin, route planning, and handover are considered. In the inter-swarm layer, the design of the routing strategy is studied. Particularly, the ISCCC scheme now can be designed by considering the UAV swarm as a virtual movable antenna system. In the swarm-terminal layer, the solutions of challenging PHY security and mobile edge computing are investigated.
    \item We demonstrate the superiority of ISCCC-enabled UAV swarms flying compared to the traditional scheme. By studying a case that minimizes the MMSE of UAV swarm flying trajectory error via ISCCC link, we can find that the ISCCC framework could realize real-time state estimation and dynamic control command updates, realizing higher trajectory accuracy compared to the UAV swarm supported by the global navigation satellite system (GNSS) system with meter-level accuracy and accumulated positioning error. Finally, the development directions of the ISCCC-enabled UAV swarm are outlined. 
\end{itemize}

\section{Designs of ISCCC for UAV Swarms Flying}

In this section, we focus on supporting UAV swarm flight based on the ISCCC strategy, detailing the corresponding design principles. Without loss of generality, we consider a UAV swarm supported by cellular networks.
\begin{figure}[t]
\centering
\includegraphics[width=2.6in]{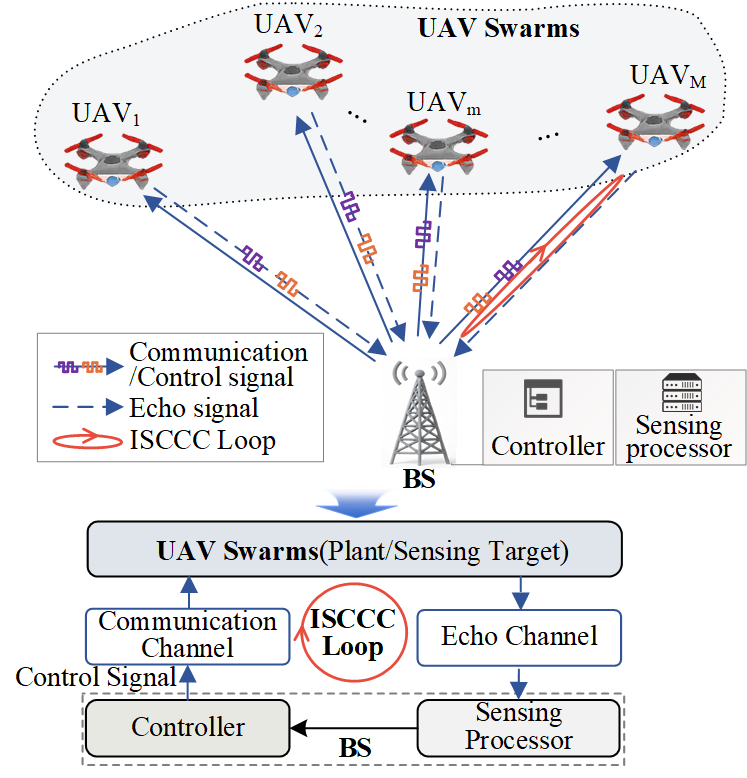}
\caption{UAV swarm flying supported by a single BS via ISCCC links.}
\label{fig:2}
\vspace{-1em}
\end{figure}
\subsection{The BS-Swarm ISCCC Links Supported by a Single BS}
As shown in Fig. \ref{fig:2}, a typical LAWN system may consist of a UAV swarm and a single BS, where the BS acts as the management center and ISCCC links are designed to support reliable and effective swarm flight.

A primary concern is how to design ISCCC protocols for UAV swarm flying, considering the interrelationship among ISCCC components and limited channel resources. Specifically, for sensing, the BS periodically transmits ISAC signals to monitor the spatial deployment of the UAV swarm. The detection accuracy depends on allocated channel resources, creating a trade-off between sensing accuracy and communication capacity. For control, the BS determines control commands based on the swarm status and transmits them via the communication link. To ensure stable control, the channel capacity for control must meet a minimum threshold. For communication, the BS may broadcast LAWN status to each UAV, while UAV swarms need to offload data to the BS. Furthermore, computing critically affects system delay. Various tasks require computation, such as decoding high-quality video streams from UAVs and processing raw sensing data at the BS, which can impact overall system efficiency.

To address these aspects, as illustrated in Fig. \ref{fig:2}, a classic mathematical model for ISCCC in UAV swarms aims to minimize the control cost (e.g., the linear quadratic regulator (LQR) cost \cite{ISCCC1}) by adjusting power allocation coefficients for the UAVs. The optimization constraints include the minimum data rate for stable control links, total power consumption, computing delay limits, sensing accuracy requirements, and specific swarm formation requirements. This model can support reliable and effective flight for UAV swarms under centralized management.

However, several issues require further exploration. First, supporting reliable ISCCC with limited wireless resources becomes challenging when the swarm contains a large number of UAVs. On one hand, next-generation multiple access (NGMA) techniques, traditionally focused on access, could be generalized to handle integrated ISCCC in UAV swarm scenarios. On the other hand, lightweight transmission schemes, e.g., semantic transmission, are needed to reduce data size. Second, designing ISCCC for realistic control tasks is demanding. Control tasks with different data rate requirements (e.g., path routing, position adjustment, transmission strategy updates) need to be classified. Correspondingly, ISCCC schemes must be designed to ensure system stability and robustness.

\subsection{The BS-Swarm ISCCC Links Supported by Multiple BSs}
\begin{figure}[!t]
\centering
\includegraphics[width=3in]{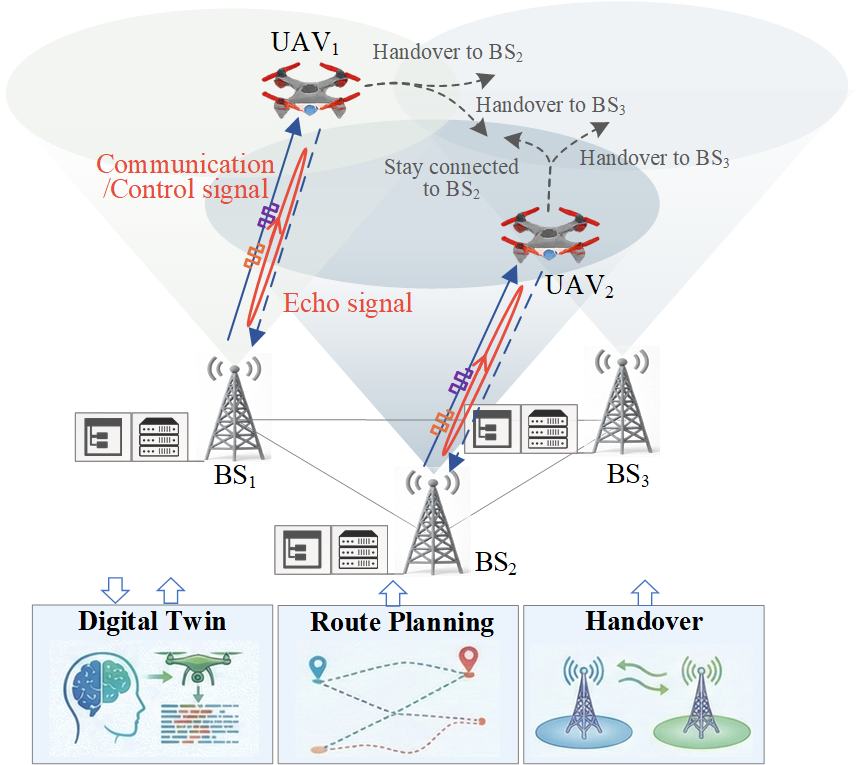}
\caption{UAV swarm flying supported by multiple BSs via ISCCC links.}
\label{fig:3}
\vspace{-1em}
\end{figure}

Given the operational scope of UAV swarms, a single BS is often insufficient due to coverage limitations. Therefore, designing reliable and effective ISCCC strategies for a UAV swarm served by multiple BSs is challenging. This involves designing multiple {\it reflex arcs} and enabling intelligent cooperation among them, as shown in Fig. \ref{fig:3}. Promising challenges and solutions are discussed below.

\subsubsection{Digital Twin}
Multiple BSs form a unified network to support UAV swarm flight. For effective network management, a digital twin platform with substantial computational power can be deployed at a central management node. This platform comprises virtual entities representing the UAV swarms, multiple BSs, and LAWN services. In operation, the digital twin platform is updated via the sensing links of the ISCCC reflex arcs. Task-relevant information is then extracted through the platform's computing module, and finally, control commands generated by the platform are transmitted via the ISCCC reflex arcs. Overall, the digital twin platform facilitates real-time monitoring and management of large-scale LAWN. Notably, its functional mechanism can be viewed as a virtual embodiment of the ISCCC reflex arc, with enhanced capability to integrate advanced AI technologies. However, constructing an efficient and sustainable digital twin platform remains challenging, requiring optimal allocation of available SCCC resources at UAV swarms, BSs, and the management center while considering service requirements and energy constraints.

\subsubsection{Route Planning}
In the near term, achieving seamless wireless coverage for LAWN based on cellular BSs is difficult. Consequently, airspace routes covered by high-quality networks are limited. Thus, planning reliable flight routes for UAV swarms under various constraints is a key challenge. Flight reliability can be jointly assessed by the risks of swarm collisions, system failures, and the autonomous flight capability of individual UAVs. Constraints include ISCCC overhead, network deployment sensing, route utilization efficiency, ground obstacles, aircraft noise, and weather conditions. Specifically, ISCCC overhead can be evaluated by control frequency, computational load, communication capacity for service data, and sensing pilot overhead. Importantly, airspace routes need to be dynamically scheduled, allocated, and released for UAV swarms. Therefore, flight routes represent a novel type of resource that must be managed in LAWN, distinct from traditional spectrum, time, space, and computational resources, yet deeply coupled with the design of ISCCC schemes.

\subsubsection{Handover}
In cellular-supported LAWN, handovers for UAV swarms will be significantly more frequent, challenging the dynamic release and allocation of channel resources. Traditionally, a handover is triggered when the signal strength difference between the source and target BSs reaches a threshold. Recent approaches can enhance this process: first, handover decisions can incorporate sensing parameters (e.g., sensing distance \cite{HO}) alongside signal strength to improve success probability. Second, BSs can operate cooperatively by sharing control and contextual information, enabling pre-triggered handovers based on combined control and geometric data. Third, computational capabilities can transform multiple BSs into intelligent agents. In a UAV-centric network mode, deployed AI models can improve handover efficiency and reliability. However, beam tracking remains an essential challenge for enabling UAV swarm handovers. Jointly ensuring beam coverage, handover reliability, and reduced network energy consumption therefore requires further research.

\section{Designs of ISCCC for UAV Swarms Functioning}
Notably, the UAV swarms will not only work under the command of BSs, but also act as management centers based on powerful abilities, such as those in the industrial production scenario. Meanwhile, compared to UAV swarm flight management, the ISCCC designs for UAV swarm operation will be more complex due to the highly dynamic deployment characteristics. Therefore, in this section, we focus on how UAV swarms operate under the ISCCC strategy.\par
\begin{figure}[t]
    \centering
    \includegraphics[width=2.8in]{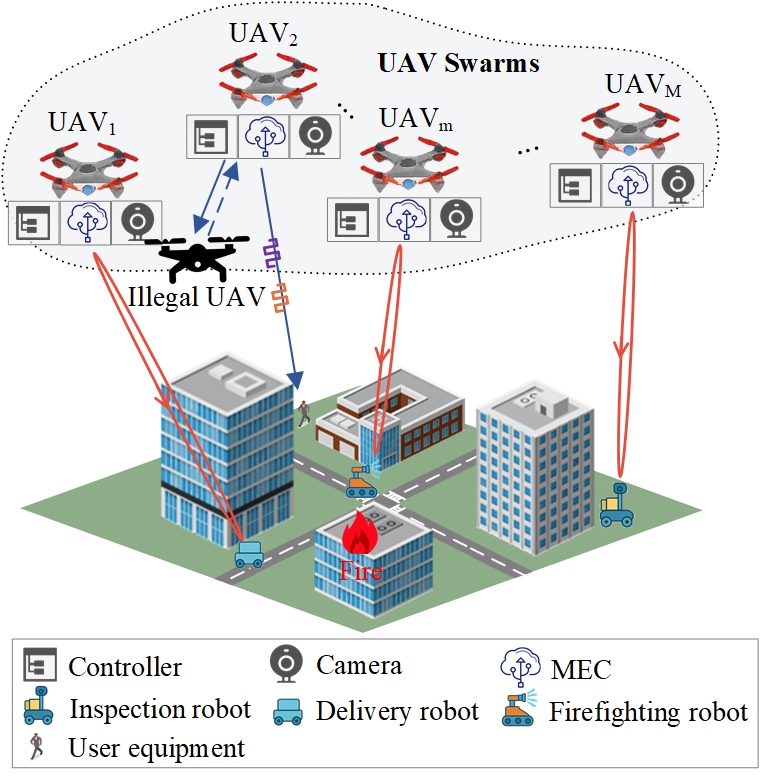}
    \caption{UAV swarm functioning for terrestrial terminals via ISCCC links.}
    \label{fig:4}
    \vspace{-1em}
\end{figure}

Taking the wild rescue scenario as an example, as shown in Fig. \ref{fig:4}, the UAV swarms could sense the situation based on the camera payload or the terrestrial monitoring devices, conduct online computation to extract the disaster-related information, control the terrestrial robots to conduct the rescue task, and always transmit business data to the terrestrial users. Notably, the ISCCC based on UAV swarms emphasizes the ability of edge computation to perceive the situation and generate control commands timely. Additionally, UAV swarms will face increased security challenges. To this end, the typical challenges and solutions are discussed below.
\subsection{Mobile Edge Computing}
Mobile Edge Computing (MEC) serves as a critical enabler for ISCCC in UAV swarms by decentralizing computational resources to the network edge. This architecture not only leverages ground-based edge servers but also incorporates UAVs as distributed computing nodes within the edge computing paradigm. In such a collaborative framework, UAVs function not merely as sensing or communication relays, but actively participate as computational entities for task processing. This approach effectively mitigates the inherent constraints of UAVs regarding onboard processing capacity and finite energy reserves. Embedded with MEC, the UAV swarm transitions from a basic execution tool to a cognitive agent that performs real-time perception, relaying, and computation. This transformation substantially augments the system-wide sensing and computing capacities of the low-altitude wireless network.

The deep integration of MEC and ISCCC is realized through three primary operational modes. First, the local integrated sensing and computation mode emphasizes lightweight processing of sensory data within the swarm itself, minimizing latency for mission-critical control tasks \cite{MEC-1}. Second, the edge-assisted intensive computation mode offloads raw or pre-processed data to ground-based or aerial edge servers to handle resource-intensive workloads such as large-scale target recognition and machine learning model training \cite{MEC-2}. Third, collaborative computation with ground users mode distributes specific computational tasks to capable ground user equipment \cite{MEC-3}, thereby optimizing the overall computational load and energy consumption of UAV swarms.\par

Nevertheless, several core challenges must be overcome to realize this integrated vision. The first challenge involves managing the complex coupling between sensing metrics and system resources. Critical trade-offs exist among sensing accuracy, sensing power, and overhead, all of which must be coordinated with communication and computational resources in real time. Furthermore, the inherent mobility of UAV swarms continuously threatens the stability of resource allocation mechanisms, rendering traditional static or quasi-static optimization approaches inadequate. Additionally, a cross-layer protocol design is essential to ensure end-to-end performance, necessitating tight coordination among physical layer resource allocation, medium access control, and network layer routing to support seamless ISCCC cooperation.

To address these challenges, future research should focus on several promising directions. Establishing a comprehensive multi-objective optimization framework is fundamental, one that can simultaneously incorporate cross-domain performance metrics such as the Fisher Information Matrix for sensing, LQR cost for control, channel capacity for communication, and computational latency for computation. For dealing with system dynamics, advanced dynamic programming and Deep Reinforcement Learning (DRL) offer powerful tools for developing real-time, adaptive decision-making algorithms. Ultimately, through the design of hierarchical task offloading and resource management strategies, it becomes feasible to achieve global and dynamic resource equilibrium within the ISCCC-enabled UAV swarm architecture.

\subsection{PHY Security}
Due to the highly open operating environment of UAV swarms, the dynamic nature of network nodes, and the broadcast characteristics of wireless communication links, the ISCCC system faces severe security threats, including eavesdropping, jamming, and spoofing \cite{HBX}. Typically, upper-layer encryption algorithms are extremely complex, making them difficult to deploy on resource-constrained UAVs. Thus, we focus on analyzing the unique characteristics of these threats in the ISCCC system and their potential solutions from a physical layer security perspective.

\emph{Anti-eavesdropping for ISCCC:}  Eavesdroppers are deployed within the coverage areas of UAV and ground nodes to illegally eavesdrop on sensitive flight data. In the ISCCC system, UAVs mainly fly at low altitudes, which means that eavesdropping links are often line-of-sight channels. This presents a huge opportunity for eavesdroppers. On the one hand, eavesdroppers can receive relatively strong legitimate signals; on the other hand, they can eavesdrop on ISCCC signals at a considerable distance. To prevent eavesdropping in ISCCC, the core idea is to degrade the quality of the eavesdropper's ``line-of-sight channels", where introducing spatial degrees of freedom is a highly promising solution. From the perspective of UAVs, they can degrade eavesdropping channels by deploying multiple antennas and combining them with beamforming techniques enabled by artificial noise. From the perspective of the UAV operator, specialized jammers can be deployed to interfere with eavesdroppers, or auxiliary equipment such as RIS can be used to alter the environment of the eavesdropping channel.

\emph{Anti-jamming for ISCCC:} Jammer transmits the jamming signals to suppress the UAVs' channels, hindering UAVs from receiving flight control commands and transmitting collected information, interfering with navigation paths, and disrupting the communication links of UAV  swarms. To mitigate the adverse effects of jamming, one path is to design anti-jamming waveforms. Here, traditional anti-jamming techniques such as frequency hopping and spread spectrum can still be used. More advanced, deploying generative artificial intelligence models on UAVs to generate anti-jamming waveforms is also a promising technology. The second path involves eliminating jamming in the received signal. Therein, the characteristics of the UAV and jamming signals should be analyzed and the jamming signal can be suppressed afterwards. The third path is to suppress jamming in the physical space. For example, deploying reconfigurable intelligent surface to change the propagation direction of jamming signals, or designing physical isolators to artificially block jamming.

\emph{Anti-spoofing for ISCCC:} Spoofing refers to attackers sending false signals to induce UAVs to make incorrect judgments. Here, we provide several typical examples of spoofing threats in UAV swarms: 1) Navigation spoofing: Attackers send fake GPS navigation signals to UAVs, causing them to deviate from their normal flight paths; 2) Control command spoofing: Attackers impersonate the control station and send false control commands to UAVs, causing illegal hijacking of UAVs; 3) Communication link spoofing: Attacks impersonate fake collaborative UAVs and communicate with other UAVs, thereby disrupting the mission execution of the UAV swarms. To prevent drones from being deceived, the key lies in identifying whether the received signal is legitimate. For example, since low-cost UAVs often have significant hardware distortion, specific emitter identification technology can be used to distinguish UAVs. Furthermore, UAVs move at high speeds, meaning their channels are rapidly time-varying. Thus, physical layer authentication via channel characteristics is also a very promising approach.

\section{Designs of ISCCC for UAV Swarms Self-Organizing}

\begin{figure}[t]
\centering
\includegraphics[width=2.8in]{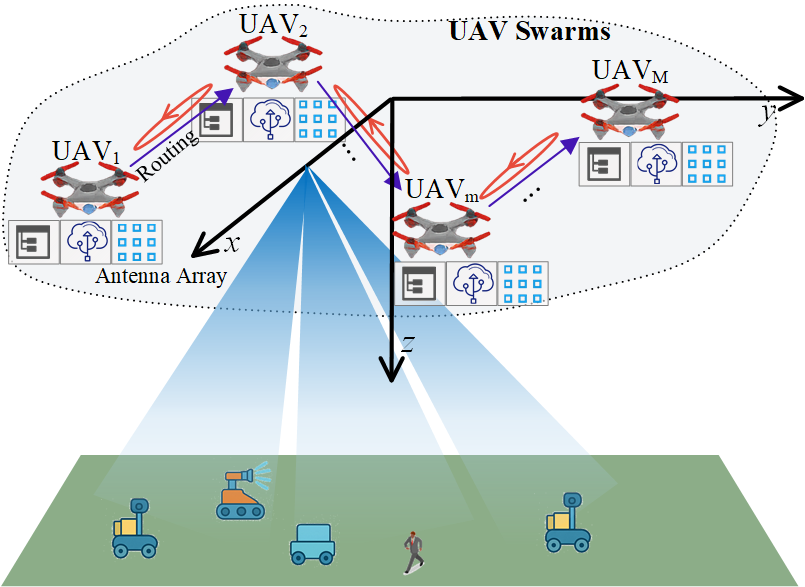}
\caption{UAV swarm self-organizing via inter-swarm ISCCC links.}
\label{fig:5}
\vspace{-1em}
\end{figure}

Furthermore, the ISCCC principle could also play a crucial role in the design of flying ad hoc networks (FANETs). Consider a UAV swarm connected via a MESH network, which is composed of one or several main UAVs and other slave UAVs. In this setup, the main UAVs are expected to act as the central controllers of the ISCCC links, where the status of the slave UAVs is monitored and controlled. As shown in Fig. \ref{fig:5}, the main challenges and potential solutions for ISCCC-based inter-swarm network design are discussed below.

\subsection{Routing Strategy}
For FANETs in UAV swarms, designing a robust and timely routing strategy under dynamic topology variations is a core challenge. In this field, geography-based routing schemes such as the greedy perimeter stateless routing (GPSR) method have been proposed. In GPSR, a UAV selects the neighbor geographically closest to the final destination as the next hop. If no such neighbor exists, i.e., when a routing void occurs, packets are routed around the void using the right-hand rule. This allows GPSR to operate without maintaining end-to-end routing state, making it highly efficient for mobile wireless networks. However, this approach can lead to suboptimal and lengthy paths when circumventing voids.

Based on the ISCCC principle, we argue that geography-based routing can be enhanced into ISCCC-aware routing schemes. On one hand, the sensing capability of the main UAVs can detect the topology of neighboring UAV swarms through ISAC waveforms or visual monitoring. Assisted by onboard computing, the main UAVs can further predict topology and geographic changes among slave UAVs. As a result, a dynamically updated planarized network graph becomes available, enabling real-time selection of optimal routing paths. In essence, ISCCC can transform routing from a passive into an active process.

On the other hand, the control capability of the main UAVs can not only synchronize routing strategies across the swarm but also adjust the formation itself to maintain highly reliable and efficient routing. For example, slave UAVs can be commanded to move to specific 3D positions based on detected or predicted geographic variations. This allows the swarm formation to become more adaptive, reducing frequent routing voids common in traditional GPSR. Thus, control capabilities hold promise for realizing swarm intelligence in complex flight scenarios.

Nevertheless, several issues must be addressed in designing ISCCC-based routing strategies. Traditional geographic routing mainly considers distance, hop count, or link lifetime. In contrast, ISCCC-based routing must holistically account for communication link quality, node computational load, real-time sensing requirements, and the priority of control commands. Moreover, ISCCC increases the dynamism of the UAV network: not only does the topology change rapidly, but so do node resource states and task loads. Routing protocols must therefore converge quickly and remain stable under such high-dimensional dynamics.

\subsection{Virtual Movable Antenna Designs}
When a UAV swarm is interconnected via ISCCC links, it can be viewed as a distributed antenna system in the sky \cite{ma}. Unlike traditional multi-antenna systems, the diameter of a UAV swarm, which is equivalent to the antenna aperture, can reach the order of kilometers. Consequently, the swarm inherently offers wide coverage and superior spatial resolution, with ISCCC transmission acting as the central nervous system that realizes these advantages. Moreover, treating each UAV as a single antenna flexibly deployed in 3D space transforms the swarm into a large-scale movable antenna system, offering far greater spatial degrees of freedom than conventional movable antenna designs.

To achieve high data rates for specific terrestrial users, the transmit/receive beamformer of the UAV swarm can be optimized. Parameters may include the control cost of UAV movement, sensing accuracy of UAV positions, swarm formation, and beamformer weights. Notably, unlike traditional trajectory optimization or movable antenna systems, UAV swarms must also account for sensing and control costs to achieve desired communication performance. Furthermore, when inter-swarm ISCCC links are coupled with air-to-ground ISCCC links, system design becomes more complex, but overall operational efficiency can be significantly improved.

Additionally, a distributed UAV swarm can achieve sensing performance comparable to synthetic aperture radar (SAR) or distributed coherent MIMO radar \cite{fanliu}. Specifically, UAV positions can be dynamically configured via inter-swarm ISCCC links to meet high-resolution radar sensing requirements. Compared to satellite-based SAR, UAV swarms offer greater deployment flexibility. Moreover, as the swarm moves in 3D space, the effective sensing aperture becomes larger and altitude-domain information can be captured. Equipped with camera payloads, the swarm can further improve sensing accuracy through multi-modal data fusion.

Nevertheless, several challenges remain. For communication, efficiently controlling swarm formation and designing corresponding codebooks are difficult and often task-dependent. For cooperative sensing, time and frequency synchronization are challenging due to cost constraints, which can severely impact sensing performance. We note that advanced AI architectures, such as large language models and federated learning, could help address these challenges, though research on AI models tailored for ISCCC remains limited.

\section{Case Study}

\begin{figure}[!ht]
\centering
\subfigure[]{
    \includegraphics[width=0.225\textwidth]{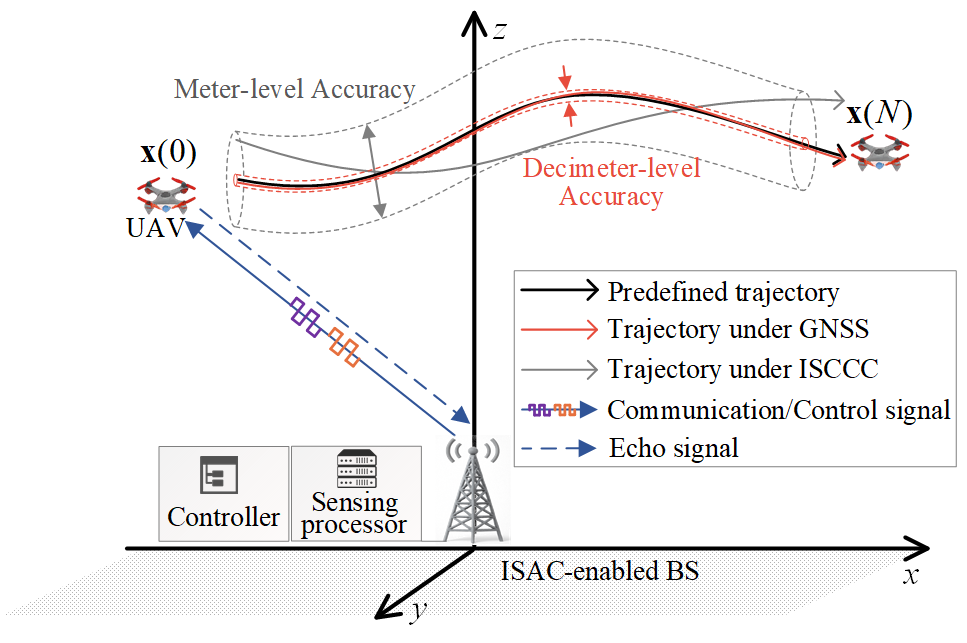}
    \label{fig:trajectory_scenario}
}
\subfigure[]{
    \includegraphics[width=0.225\textwidth]{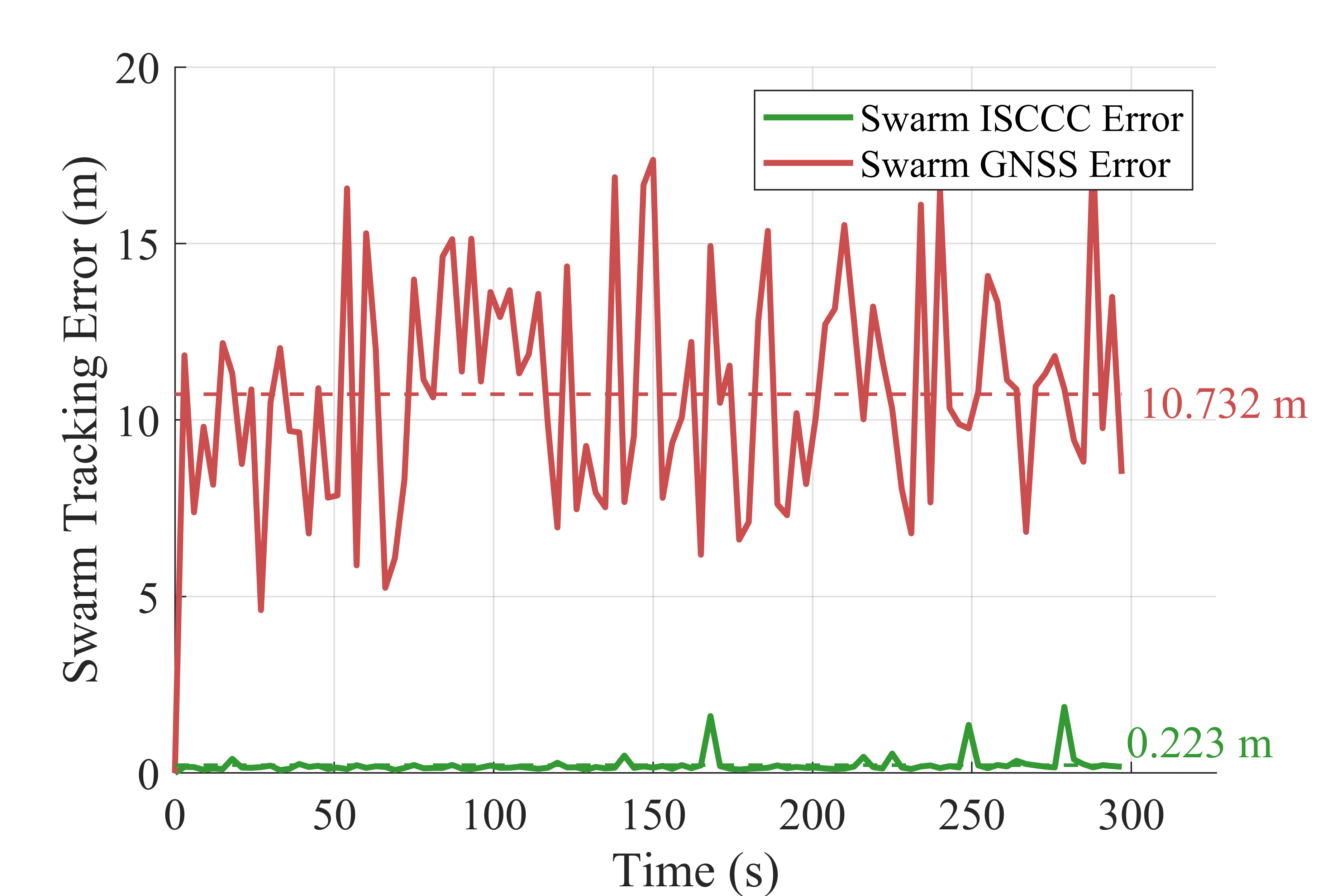}
    \label{fig:tracking_error}
}
\subfigure[]{
    \includegraphics[width=0.48\textwidth]{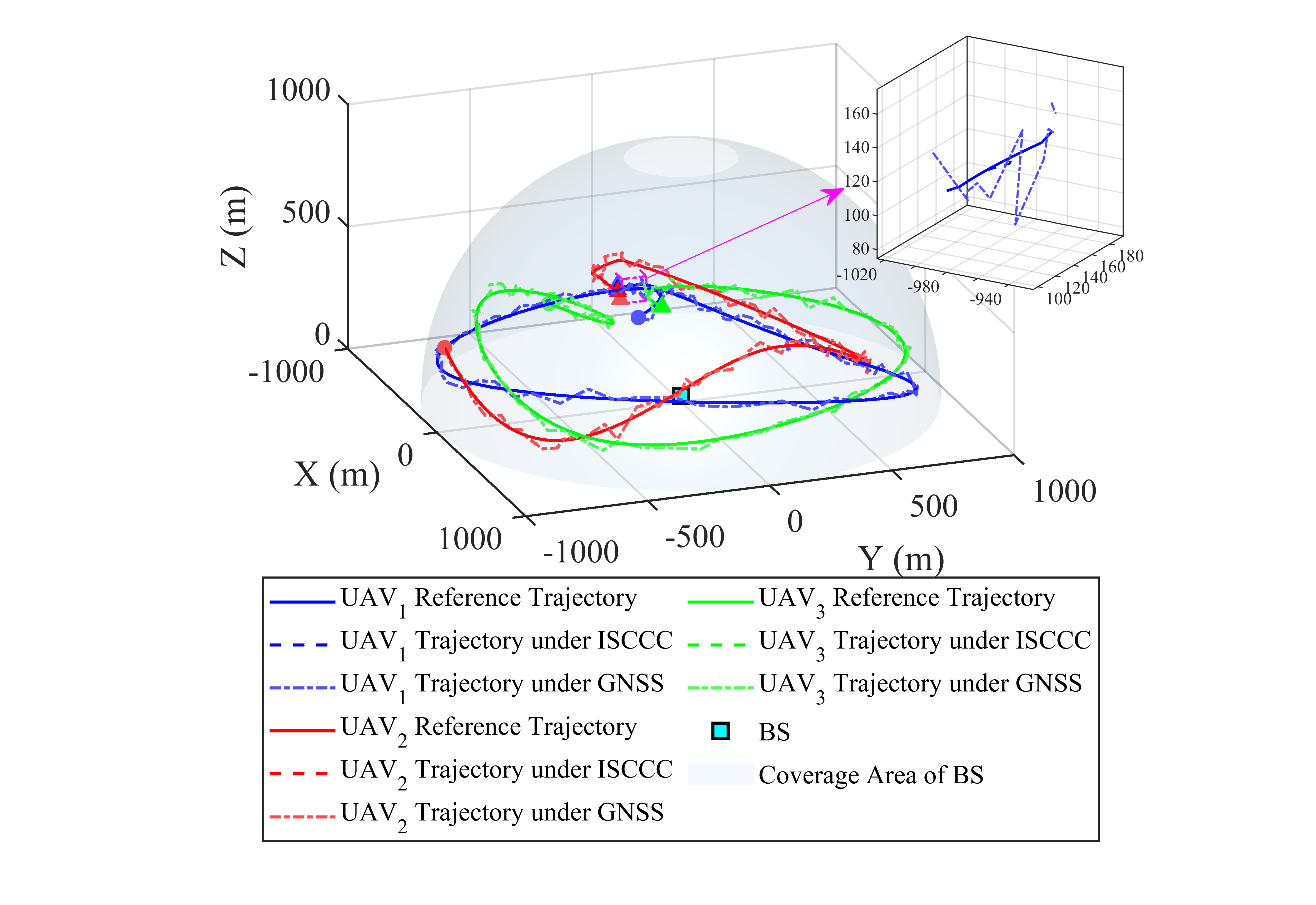}
    \label{fig:tracking_comparison}
}

\caption{Integrated visualization of UAV swarm trajectory tracking performance enabled by ISCCC: a) ISCCC-enabled high-precision UAV swarm three-dimensional trajectory scenario, b) Average trajectory error comparison of the UAV swarm under GNSS and under ISCCC, and c) UAV swarm trajectory comparison under GNSS and under ISCCC.  
}
\label{fig:ISCCC_overview}
\vspace{-1em}
\end{figure}

To evaluate the effectiveness of the proposed ISCCC architecture in dense low-altitude UAV operations, we present a case study based on high-precision trajectory tracking. As shown in Fig.~\ref{fig:trajectory_scenario}, the BS supports a UAV swarm flying along a predefined three-dimensional path. In such a dense swarm scenario, the limitations of conventional navigation mechanisms become particularly prominent. A typical GNSS provides only meter-level positioning accuracy and operates in an open-loop manner without network-side feedback, causing its localization errors to accumulate over time without timely correction. For UAVs maneuvering at high speed within constrained airspace, such accumulated errors not only degrade path-following performance but also significantly increase the risk of airspace conflicts in LAWN.

To address this issue, we establish a continuously operating ISCCC closed loop between each UAV and the BS, serving as a ``digital reflex arc'' for swarm flight. Unlike conventional architectures that separate communication and sensing functionalities, the BS in ISCCC acts as a unified ISAC--computing--control node: the downlink ISAC waveform simultaneously performs UAV state sensing and environmental probing; the edge computing module estimates the UAVs' three-dimensional positions in real time based on the echoes; and the controller generates motion-control commands according to the latest state, which are then delivered to the UAVs through a finite-blocklength link that shares spectrum with communication services. With this millisecond-level closed-loop update, ISCCC effectively suppresses state estimation noise and promptly compensates for UAV yaw perturbations, attitude variations, and environmental disturbances, thereby achieving a level of trajectory-keeping accuracy that is unattainable under traditional GNSS open-loop navigation.

The performance difference between ISCCC and GNSS is clearly reflected in the trajectory comparison shown in Fig.~\ref{fig:tracking_comparison}. UAVs relying on GNSS exhibit substantial deviations in high-dynamic regions such as sharp turns and altitude transitions, making consistent path tracking difficult. In contrast, under the ISCCC framework, the actual UAV trajectories closely follow the intended path and demonstrate significantly smoother and more stable flight behavior. Quantitative results in Fig.~\ref{fig:tracking_error} further confirm this trend: the average trajectory error under GNSS reaches approximately 10.7~m, whereas ISCCC reduces this error to only 0.22~m, achieving a 98\% improvement. This shift from open-loop to closed-loop control greatly enhances the stability and controllability of UAV swarm flight.

\section{Future Directions}
Based on the above discussions, in this section, we present future directions for designing ISCCC-enabled UAV swarms.

\subsection{Aerial Embodied Intelligence}
Aerial embodied intelligence represents a fundamental reconceptualization of ISCCC-enabled UAV swarms. Within the embodied intelligence framework, the swarm deeply integrates physical maneuvers with information processing through tightly-coupled ISCCC reflex arcs. Future research should prioritize the development of embodied intelligence architectures that exploit the synergistic relationship between physical presence and information processing. Promising directions include constructing accurate world models capable of predicting how physical actions propagate through the ISCCC control loop, formulating unified reward mechanisms that reconcile competing objectives across SCCC domains, and establishing distributed cognitive frameworks that support emergent cooperative behaviors. These advancements will ultimately transform UAV swarms into autonomous intelligent systems capable of complex mission execution in dynamically evolving environments.

\subsection{Aerial Movable Antenna System}
When integrated with UAV swarms, the movable antenna technology effectively creates a distributed massive MIMO system with exceptional spatial degrees of freedom, allowing the swarm to actively form favorable line-of-sight links, mitigate interference, and enhance spatial multiplexing gains. However, realizing this potential requires addressing significant cross-domain challenges, particularly the joint optimization of antenna positioning, UAV trajectory, resource allocation, and task scheduling. Future research should focus on developing intelligent optimization frameworks that can manage the complex trade-offs among sensing accuracy, communication rate, computational efficiency, and control stability. Promising directions include DRL for real-time decision-making, specialized codebook designs for swarm configurations, and innovative synchronization mechanisms for cooperative sensing applications.

\subsection{Channel Modeling and Waveform Design}
The conventional channel models are insufficient to capture the cross-domain dynamic coupling inherent in ISCCC. Future channel modeling should incorporate a spatio-temporal–semantic joint representation, which not only describes physical propagation laws but also integrates task characteristics, control behaviors, and environmental semantics. Such semantic channel models can guide beamforming, power, and spectrum allocation at the physical layer, while simultaneously supporting high-level control strategies and computation task scheduling, thereby forming a data foundation for cooperative UAV swarm operation. Furthermore, future waveforms should not only maximize transmission rate or sensing resolution but also achieve a unified optimization among control stability, system latency, and energy efficiency \cite{myy1,myy2}. Moreover, AI-generated waveform design is an emerging direction where deep or generative learning models can produce adaptive waveforms based on UAV swarm states, channel semantics, or predicted environmental variations. Such intelligent waveform generation will enable highly efficient, secure, and robust ISCCC transmission.
\section{Conclusion}

In this article, we investigate the possibilities and challenges of ISCCC-enabled UAV swarms in LAWN. We first propose a three-layer ISCCC construction for UAV swarms, where the principles of designing ISCCC are different. Second, we present the core challenges and key techniques for supporting UAV swarm flying, functioning, and self-organizing via ISCCC links, including handover, MEC, and routing, etc. Third, as a case study, we show that the UAV swarm with ISCCC could achieve higher trajectory accuracy compared to the traditional GNSS-based swarm flying. Finally, we discuss potential future research directions of ISCCC-enabled UAV swarms. We emphasize that the ISCCC is promising to turn UAV swarm into an aerial robot system and aerial movable antenna systems, which could promote the development of LAWN applications.
\ifCLASSOPTIONcaptionsoff
  \newpage
\fi

\end{document}